\documentstyle[aps,prl,twocolumn]{revtex}

\begin{document}

\draft
\title{Topological approach to Luttinger's theorem
and the Fermi surface of a Kondo lattice}

\author{Masaki Oshikawa}

\address{
Department of Physics, Tokyo Institute of Technology,
Oh-okayama, Meguro-ku, Tokyo 152-8551, Japan
}

\date{September 22, 1999}

\maketitle

\begin{abstract}
A non-perturbative proof of Luttinger's theorem, based on
a topological argument, is given
for Fermi liquids in arbitrary dimensions.
Application to the Kondo lattice shows that
even the completely localized spins do contribute to
the Fermi sea volume as electrons,
whenever the system can be described as a Fermi liquid. 
\end{abstract}


Landau's Fermi liquid theory is
among the most important theories in quantum many-body problem.
At zero temperature, a Fermi liquid has a Fermi surface,
similarly to the noninteracting fermions.
One of the most fundamental results on the Fermi liquid is
Luttinger's theorem, which states that the volume inside the Fermi
surface is invariant by the interaction, if the number of particles
is held fixed.
Luttinger argued, in his 1960 paper~\cite{Luttinger},
the correction to the volume vanishes order by order
in the perturbation expansion.

Recently, there have been renewed interests in the
Luttinger's theorem.
Since Luttinger's original proof was based on the perturbation theory,
Luttinger's theorem could be violated by non-perturbative effects.
In fact, several claims of possible breakdown of Luttinger's theorem
have been reported
recently~\cite{Langer,Chub,Putikka,Sherman,Groeber}.
On the other hand, such non-perturbative effects can violate the
Fermi liquid theory itself.
In fact, the Fermi liquid theory
is known to be invalid generally in one dimension,
where Tomonaga-Luttinger (TL) liquid is the generic behavior.
Although not being a Fermi liquid,
a TL liquid in one dimension has a well-defined
Fermi surface (actually Fermi points in one dimension).
Thus the question of the validity of Luttinger's ``theorem''
still exists in this case, where Luttinger's original proof
certainly does not apply.
This question was answered recently
by a perturbative proof~\cite{Blagoev}
and a more general non-perturbative proof~\cite{YOA},
which can be applied to one-dimensional TL liquids.
However, the question on higher (especially two) dimensions
remains unanswered.
In fact, it is not clear whether a Fermi liquid
which violates Luttinger's theorem can exist.

Another interesting problem, which is not answered by the Luttinger's
perturbative proof, is the Fermi surface of the Kondo lattice.
The Kondo lattice contains a
periodic array of localized spins which are coupled to conduction electrons.
The Kondo lattice is believed to belong to the Fermi liquid
(or TL liquid, in one dimension) in some region of the
phase diagram.
Even if we assume Luttinger's theorem to be valid,
there is a problem in how to count number of particles.
It is rather difficult, by conventional methods, to
clarify whether a localized spin should be counted
as an electron (``large Fermi surface'' picture) or
not (``small Fermi surface'' picture).
In one dimension, the non-perturbative proof
of the Luttinger's theorem was also applied
to the Kondo lattice~\cite{YOA},
to show that the localized spins do participate in
the Fermi sea. (See also~\cite{Moukouri,Shibata} for numerical evidences.) 
On the other hand, there has been no definite answer
for higher dimensions, although there
are several results~\cite{Taillefer,Eder}
supporting the ``large Fermi surface'' picture.

The argument in Ref.~\cite{YOA} is a generalization of
Lieb-Schultz-Mattis (LSM) theorem~\cite{LSM}, which
was given at about the same time as
the apparently unrelated Luttinger's theorem~\cite{Luttinger}. 
Since the LSM argument itself cannot be applied to higher dimensions,
the discussion in Ref.~\cite{YOA} was restricted
to one dimension.
However, very recently the LSM argument was combined with
Laughlin's gauge invariance argument~\cite{Laughlin}
on the Quantum Hall Effect (QHE) and extended to higher
dimensions~\cite{Oshikawa}.
Inspired by this observation,
we will extend the non-perturbative
proof of the Luttinger's theorem to arbitrary dimensions
in the present letter.

We consider an interacting fermion system on a
$D$-dimensional lattice with periodic boundary conditions.
We start from a finite system of size
$L_x \times L_y \times \ldots \times L_D$, where the length
is defined so that the unit cell has the size
$1 \times 1 \times \ldots \times 1$.
The number of fermions is assumed to be conserved.
If the system satisfies a commensurability condition,
it can have a finite excitation gap~\cite{Oshikawa}.
In this letter, we will rather focus on the gapless case,
which is expected for general incommensurate
particle density.
For simplicity, let us first start with the case of
spinless fermion of single species.
We introduce a fictitious electric charge $e$ for 
each particle, and a coupling to
an externally controlled fictitious electromagnetic field.
Because of the periodic boundary condition, the
system is topologically equivalent to torus.
Following Refs.~\cite{Laughlin,Oshikawa},
we consider an adiabatic increase of a (fictitious)
magnetic flux $\Phi$ piercing through the ``hole'' of the torus
so that the uniform electric field
is induced, say, in the $x$-direction.

While in general the Hamiltonian of the system $H(\Phi)$ depends on the
flux $\Phi$ reflecting the Aharanov-Bohm (AB) effect,
the AB effect is absent when the flux reaches the unit flux quantum
$\Phi_0 = hc/e$.
We consider the adiabatic increase of the flux from $\Phi=0$
to $\Phi=\Phi_0$.
In the following, we will consider how the total momentum
of the system is changed during the adiabatic process
in two different ways, and compare those results.
In the remainder of this letter,
we take the units in which $\hbar=1$, for simplicity.

First, we analyze the momentum change in a system
of interacting fermions for general.
We remind the reader that the momentum itself is
a {\em gauge dependent} quantity in the presence of the
gauge field; a meaningful comparison between momenta can only be made
under the same gauge choice.
In a simplest gauge choice, the AB flux $\Phi$ is represented by
the uniform vector potential $A_x = \Phi /L_x$ in the $x$-direction.
In this gauge,
the Hamiltonian always commutes with the
translation operator $T_x$ to the $x$-direction.
We further assume that the translation symmetry is not
spontaneously broken, as it should not in a Fermi liquid.
Thus the ground state is an eigenstate of the total momentum $P_x$:
$ P_x | \Psi_0 \rangle = P_x^0 | \Psi_0 \rangle$ with
the eigenvalue $P_x^0$.
The $x$-component of the total momentum $P_x$ is related to
$T_x$ as $T_x = e^{i P_x}$.
After the adiabatic process,
the original groundstate $|\Psi_0 \rangle$ evolves into
some state $| \Psi_0' \rangle$.
While the state $|\Psi_0'\rangle$ could be different
from $|\Psi_0 \rangle$,
it belongs to the same eigenvalue $P_x^0$ of $P_x$,
because the Hamiltonian always commutes with $T_x$ (and thus $P_x$)
in the uniform gauge~\cite{Oshikawa} during the adiabatic process.
Although it naively means that the momentum is unchanged
after the adiabatic process, it is not true.
The Hamiltonian $H(\Phi_0)$ with the unit flux quantum
in the uniform gauge is different from the original one $H(0)$,
although the spectrum should be identical.
Namely, they correspond to different choices of the gauge
for the same physics.
In order to get back to the original gauge,
we must perform a large gauge transformation~\cite{Oshikawa}
\begin{equation}
U = \exp{[ \frac{2 \pi i}{L_x} \sum_{\vec{r}} x n_{\vec{r}}]},
\end{equation}
where $n_{\vec{r}}$ is the particle number operator
at site $\vec{r}$, and $x$ is the $x$-coordinate of $\vec{r}$.
This transforms the Hamiltonian $H(\Phi_0)$ back to the
original one: $U H(\Phi_0) U^{-1} = H(0)$.
After this gauge transformation, the adiabatic evolution of the
groundstate becomes $U | \Psi_0' \rangle$.

Now we can examine the total momentum $P_x$ of this state,
and compare it with the original one $P_x^0$.
Here we can employ the arguments used in the LSM theorem
and its generalizations~\cite{LSM,YOA}.
By using the identity
\begin{equation}
U^{-1} T_x U = T_x \exp{[ 2 \pi i \sum_{\vec{r}} \frac{n_{\vec{r}}}{L_x} ]}
\end{equation}
we see that $U | \Psi'_0 \rangle$ is an eigenstate of $P_x$ with
\begin{equation}
P_x = P_x^0 + 2 \pi \nu L_y L_z \ldots L_D,
\label{eq:pxshift2}
\end{equation}
where $\nu$ is the particle density (number of particles per unit cell).
This result is valid regardless of the interaction strength.

Next, we analyze the momentum change assuming that the system
is a Fermi liquid.
The Fermi liquid is described in terms of quasiparticles, which are
almost non-interacting.
More precisely, the low-energy
effective Hamiltonian of a Fermi liquid is given by
\begin{equation}
{\cal H} \sim
\sum_{\vec{k}} \epsilon(\vec{k}) \tilde{n}_{\vec{k}}
+ \sum_{\vec{k},\vec{k}'}
f(\vec{k},\vec{k}') \tilde{n}_{\vec{k}} \tilde{n}_{\vec{k}'},
\label{eq:flham}
\end{equation}
where $\tilde{n}_{\vec{k}}$ is the {\em quasiparticle} number operator
of momentum $\vec{k}$.
Namely, there is an interaction energy due to the second term
but no scattering between the quasiparticles.
Thus the eigenstates of $\tilde{n}_{\vec{k}}$ are also
eigenstates of Hamiltonian.
In the groundstate, the Fermi sea (region inside the Fermi surface)
is completely filled with quasiparticles, while the outside is empty
in terms of quasiparticles.
Excitations on the groundstate are given by quasiparticles
outside the Fermi sea and/or quasiholes inside the Fermi sea.
In fact, the quasiparticle (or quasiholes)
are free from scattering only in the vicinity
of the Fermi surface;
the very notion of quasiparticle/hole is useful only in this case.
The Fermi liquid theory is valid for the low-energy phenomena,
in which the relevant excitations consist only of quasiparticles
(quasiholes) near the Fermi surface.

Let us define the Fermi sea volume $V_F^{(L)}$
in the finite size system $L_x \times L_y \times \ldots \times L_D$.
The quasiparticles are scattering free, and
their momenta are discretized as in the case of free particles.
Thus we can define the Fermi sea volume $V_F^{(L)}$
by an integer ``occupation number'' of the quasiparticles $N_F^{(L)}$:
\begin{equation}
	V_F^{(L)} = \frac{(2 \pi)^D N_F^{(L)}}{L_x L_y \ldots L_D}.
\label{eq:vfl}
\end{equation}
Although the quasiparticles are not
free from scattering (and thus are not meaningful)
away from the Fermi surface,
this expression is still valid because the Fermi sea volume is
uniquely determined by its surface.
The $V_F^{(L)}$ should
approach the true volume of the Fermi sea $V_F$, in the
thermodynamic limit $L_j \rightarrow \infty$.

The adiabatic evolution is determined by the low-energy
effective Hamiltonian~(\ref{eq:flham}).
In the Fermi liquid theory, the charge of the quasiparticle
is identical to that of the original particle $e$.
The coupling of the quasiparticles to the uniform vector potential
$A_x$ is thus given by the substitution of the momentum
$k_x \rightarrow k_x + e A_x /c$ in the Hamiltonian.
After the adiabatic insertion of the unit flux quantum,
and getting back to the original Hamiltonian by
the gauge transformation,
each quasiparticle gets a momentum shift:
$k_x$ is increased by $2 \pi /L_x$.
This produces quasiparticles on one side of the Fermi surface,
and quasiholes on the opposite side.

Since the result of the adiabatic process is equivalent to
the shift of the whole Fermi sea by $2\pi/L_x$,
the change of the $x$-component
of total momentum $P_x$ of the system during
the adiabatic process is given by
\begin{equation}
\Delta P_x = \frac{2 \pi}{L_x} N_F^{(L)}
\label{eq:pxshift1}
\end{equation}
We note that the only changes after the adiabatic process
involve the quasiparticles and quasiholes near the Fermi surface,
so that the Fermi liquid theory is still valid.
To violate eq.~(\ref{eq:pxshift1}), the system must break some of the
properties of Fermi liquid used in the present argument.
For example, if a quasiparticle had a charge $e'$ which is different
from the charge $e$ of the original particle,
we would obtain a different result.

Now, comparing the two results eqs.~(\ref{eq:pxshift2}) and
(\ref{eq:pxshift1}) obtained with different arguments,
we obtain
$ N_F^{(L)}/ L_x  - \nu L_y L_z \ldots L_D = \mbox{(integer)} $,
where we have used the fact that
each component of momenta is defined modulo $2 \pi$.
Let us choose the system size so that $L_x, L_y, \ldots$
and $L_D$ are mutually prime with the others. 
We also assume $L_x = q l_x$ where $l_x$ is an integer.
(It should be recalled that system size should be an integral
multiple of $q$, to allow the filling factor $\nu = p/q$.)
Then, from eq.~(\ref{eq:vfl}) we obtain
$N_F^{(L)} - p l_x L_y L_z \ldots L_D =  L_x \times \mbox{(integer)}$.

Furthermore, we can consider other adiabatic processes,
in which the gauge field is induced in one of the other directions
$y, z, \ldots$, instead of $x$.
Similar calculations for these cases lead to
$N_F^{(L)} =  L_{\alpha} \times \mbox{(integer)}$,
where $\alpha = y, z, \ldots, D$.
Because we have chosen the lengths $L_j$'s mutually prime,
we conclude that
$N_F^{(L)} - p l_x L_y L_z \ldots L_D = n L_x L_y L_z \ldots L_D$,
where $n$ is an integer.
Writing this in terms of Fermi sea volume, we arrive at
\begin{equation}
\frac{V_F}{(2 \pi)^D} - \nu = n,
\label{eq:theorem}
\end{equation}
where we have replaced the Fermi sea volume $V_F^{(L)}$
for the finite size system by its thermodynamic limit $V_F$,
because this relation is exact already for the finite system.
The thermodynamic limit $V_F$ should be independent of
our special (mutually prime) choice of $L_j$'s,
if $V_F$ is well-defined.

The relation~(\ref{eq:theorem}) is nothing but the statement of
Luttinger's theorem.
The integer $n$ corresponds to the number of completely filled bands.
It is valid also when the Fermi sea consists of several disjoint
regions, if $V_F$ is understood as the sum of volumes of
all regions.
Our proof is much simpler than the original one~\cite{Luttinger}.
Moreover, in contrast to Ref.~\cite{Luttinger},
our argument is non-perturbative and relies only
on some of the basic properties of Fermi liquid.

It is straightforward to extend our argument to spinful electrons.
When the numbers of up-spin electrons and down-spin electrons are
conserved separately, we consider the fictitious electromagnetic
field coupled to only up-spin (or down-spin) electrons.
Assuming the spinful Fermi liquids,
the volume of the Fermi sea $V_F^{\sigma}$ for spin $\sigma$ is given by
$V_F^{\sigma} = (2 \pi)^D \nu_{\sigma}$,
where $\nu_{\sigma}$ is the number of particles with spin $\sigma$
per unit cell.
For the spin-symmetric case $\nu_{\uparrow} = \nu_{\downarrow}$,
it reads $V_F = V_F^{\uparrow} = V_F^{\downarrow} = (2 \pi)^D \nu/2$
where $\nu$ is the total particle density
$\nu_{\uparrow} + \nu_{\downarrow}$.

As a nontrivial application, let us consider the Kondo lattice.
Luttinger's original perturbative proof does not apply to this
case, and the question on the volume of the Fermi sea has remained.
For the sake of clarity, we consider the Kondo lattice model
given by the Hamiltonian
\begin{equation}
H = - \sum_{j,k} t_{jk} c^{\dagger}_{j \sigma} c_{k \sigma} + \mbox{h.c.}
+ \sum_j U_j  c^{\dagger}_{j \uparrow} c_{j \uparrow}
		c^{\dagger}_{j \downarrow} c_{j \downarrow}
		+ \sum_l J_l \vec{s}_l \cdot \vec{S}_l,
\end{equation}
where $c^{\dagger}_{j \sigma}$ and $c_{j \sigma}$ are standard
Fermion creation/annihilation operators at site $j$ with spin $\sigma$,
$\vec{s}_l = c^{\dagger}_{l \alpha} \vec{\sigma}^{\alpha \beta} c_{l \beta}/2$
is the spin operator of the conduction electron,
and $\vec{S}_l$ is the localized spin at site $l$.
As in the previous case, we couple the fictitious electromagnetic field
only to the up-spin electrons.
After the adiabatic insertion of the AB flux of unit flux quantum,
we make the gauge transformation as in the previous cases.
However, the naive one
$ U_{\uparrow}^e =
\exp{[ \frac{2 \pi i}{L_x} \sum_{\vec{r}} x n_{\vec{r} \sigma}]}$,
does not bring the Hamiltonian back to the original one,
because it changes the Kondo coupling.
In order to recover the original Hamiltonian,
we must also twist the localized spins.
The transformation 
\begin{equation}
U_{\uparrow} =
\exp{[ \frac{2 \pi i}{L_x}
\sum_{\vec{r}} x ( n_{\vec{r} \sigma} + S^z_{\vec{r}})]}
\end{equation}
does the required job.
We obtain the total momentum after the adiabatic process as
\begin{equation}
P_x = P_x^0 + 2 \pi [ \nu_{\uparrow} + N_s (S + m) ]  L_y L_z \ldots L_D,
\end{equation}
where $N_s$ is the number of localized spins per unit cell and
$m$ is the magnetization per single localized spin.
The special contribution proportional to $S$ comes
from the boundary term $\exp{(2 \pi i N_s S^z_1)}$ appearing in
$U_{\uparrow}^{-1} T_x U_{\uparrow}$,
similarly to the one dimensional case~\cite{LSM,OYA,YOA}.

Thus, provided that the system belongs to a spinful Fermi liquid,
the volume of the Fermi sea is given by
$V_F^{\sigma} = (2 \pi)^D [ \nu_{\sigma} +  N_s (S \pm m) ]$,
where $\pm$ takes $+$ for $\sigma = \uparrow$
and $-$ for $\sigma = \downarrow$.
For the spin-symmetric case $\nu_{\uparrow} = \nu_{\downarrow}$
and $m=0$, we obtain
\begin{equation}
V_F = V_F^{\uparrow} = V_F^{\downarrow}
= \frac{(2 \pi)^D}{2} [ \nu + 2 N_s S ],
\end{equation}
for the total particle density
$\nu = \nu_{\uparrow} + \nu_{\downarrow} = 2 \nu_{\uparrow}$.
This is exactly what we obtain if we apply the Luttinger's theorem
to the Anderson-type model in which the localized spins are
represented by electrons.
It means that the localized spin $S$ does contribute to the Fermi
sea volume as $2S$ electrons, even though it is completely
immobile.
This is the picture conventionally
called as the ``large Fermi surface''.

It should be noted that we did {\em not} answer the non-trivial
question whether (or when) the Kondo lattice belongs to the Fermi liquid.
We have only proved that, if the Kondo lattice is a Fermi liquid
(as it is believed to be true in some region of the phase diagram),
the localized spins participate in the Fermi sea.

Finally, let us comment on
claims~\cite{Langer,Chub,Putikka,Sherman,Groeber}
of the violation of the Luttinger's theorem.
There are several possibilities regarding the apparent contradiction
to our non-perturbative proof.
Of course, it should be checked whether our
argument applies to the model under consideration.
However, our argument does apply to a very wide range of lattice models,
including the Hubbard and $t$-$J$ models, for which the violation
of the Luttinger's theorem has been also proposed.
A possibility is that the system is not a Fermi liquid in these cases.
In other words, a violation of the Luttinger's theorem requires
the system to be a non-Fermi liquid.
We note that, however, merely not being a Fermi liquid is insufficient,
as the TL liquid in one dimension does satisfy the Luttinger's
theorem~\cite{Blagoev,YOA}.
Our approach could be extended to a non-Fermi liquid which has
an appropriately defined Fermi surface,
if such a liquid does exist.
Our argument reveals a rigid relationship between
the structure of low-energy excitations and the Fermi sea volume.

Another possibility is that the claimed violation
of Luttinger's theorem is actually incorrect.
In particular, numerical results are only available for
restricted system size and/or temperature, and can miss the
possibly small singularity at the true Fermi surface.
On the other hand, even if they are incorrect in identifying the
true Fermi surface, they might still be of physical relevance because
the actual experiments are done also at finite energy scale;
the experimentally measured ``Fermi surface'' could be different
from the true Fermi surface defined in the low-energy limit,
to which our argument applies.
In any case, our definite result on the Fermi surface 
of the Fermi liquid in the low-energy limit would be useful as a guideline.
Claims of the violation of the Luttinger's theorem should be examined
in the light of the present result.

\bigskip

During the forty years after the Luttinger's paper~\cite{Luttinger},
several examples of ``quantization'' of a physical
quantity have been found in many-body physics.
Namely, despite the complexity of the interacting many-body states,
some physical quantity takes a special value which is
stable against various perturbations such as interaction strength.
Presumably the most natural understanding of such a quantization is
given by a topological argument.
Indeed, typical examples of the quantization -- QHE
and the quantized magnetization plateaus have been related
to topological mechanisms~\cite{Laughlin,OYA,Oshikawa}.

Luttinger's theorem perhaps does not look like a quantization,
because the volume of the Fermi sea takes continuous values
depending on the particle density.
However, the insensitivity to the interaction resembles to
other quantization phenomena, and may well be regarded
as a certain kind of quantization,
especially when written as in eq.~(\ref{eq:theorem}).
In fact, we have revealed a close theoretical relationship among
Luttinger's theorem, QHE and magnetization plateaus. 
In addition, our argument can be related also
to the chiral anomaly in quantum
field theory~\cite{Jackiw}.
Luttinger's theorem might be actually the first example of
the topological quantization discovered in quantum many-body problem,
although the topological understanding has been missing for a long time.

\bigskip

I would like to thank Ian Affleck, Hal Tasaki and Masanori Yamanaka
for stimulating discussions which were essential for the present work.
I am also grateful to an anonymous referee for pointing out a
logical gap in the original manuscript.
This work is supported by Grant-in-Aid from
Ministry of Education, Science, Sports and Culture of Japan.

\end{document}